Tartaric acid in red wine as one of the key factors to induce superconductivity in FeTe$_{0.8}$S$_{0.2}$

Keita Deguchi[1,2,3], Tohru Okuda[4], Yasuna Kawasaki[1,2,3], Hiroshi Hara[1], Satoshi Demura[1,2,3], Tohru Watanabe[1,2], Hiroyuki Okazaki[1,3], Toshinori Ozaki[1,3], Takahide Yamaguchi[1,3], Hiroyuki Takeya[1,3], Fumie Saito[4], Masashi Hisamoto[4], and Yoshihiko Takano[1,2,3]

[1] National Institute for Materials Science, 1-2-1, Sengen, Tsukuba, 305-0047, Japan

[2] Graduate School of Pure and Applied Sciences, University of Tsukuba, 1-1-1 Tennodai, Tsukuba, 305-8571, Japan

[3] Japan Science and Technology Agency - Transformative Research-project on Iron-Pnictides, 1-2-1, Sengen, Tsukuba, 305-0047, Japan

[4] The Institute of Enology and Viticulture, University of Yamanashi, 13-1 Kitashin-1-chome, Kofu, 400-0005, JAPAN

Corresponding author: Keita DEGUCHI

E-mail address: DEGUCHI.Keita@nims.go.jp

Address: 1-2-1 Sengen, Tsukuba 305-0047, Japan

Phone: +81-29-859-2842

Fax: +81-29-859-2601

Abstract

The red wine dependence of superconductivity in FeTe$_{0.8}$S$_{0.2}$ was investigated.

Samples with a higher shielding volume fraction had a tendency to show a higher concentration of tartaric acid in red wine. We found the tartaric acid is one of the key factors to induce superconductivity in $FeTe_{0.8}S_{0.2}$.

Introduction

Since the discovery of a new superconductor $LaFeAsO_{1-x}F_x$ [1], various iron-based superconductors have been discovered [2–4]. Among them, FeTe, one of the parent compounds of the Fe-based superconductors, has the simplest crystal structure. FeTe exhibits antiferromagnetic ordering around 70 K and does not show superconductivity. The substitution of S for Te sites suppresses the antiferromagnetic order and induces superconductivity [5]. Using a solid-state reaction, as-grown samples of $FeTe_{1-x}S_x$ do not show superconductivity although the antiferromagnetic ordering seems to be suppressed. So far, we have found that zero resistivity in $FeTe_{0.8}S_{0.2}$ is induced by exposure to air, water immersion and oxygen annealing [6-8]. We also have discovered that the superconductivity in a $FeTe_{0.8}S_{0.2}$ is induced by soaking the sample in alcoholic beverages [9]. Red wine, white wine, beer, Japanese sake, whisky, shochu are more effective in inducing superconductivity in $FeTe_{0.8}S_{0.2}$ than water. The most effective liquid among the alcoholic beverages is red wine. The shielding volume fraction of a sample heated in red wine is more than 6 times larger than that of the samples heated in water. Therefore, we assume that some components in red wine are an important factor in inducing superconductivity in $FeTe_{1-x}S_x$. An investigation of these effective components will provide us with a novel route for discovering new superconductors. We prepared 6 types of red wine made from different grapes in order to investigate the red wine dependence of superconductivity in $FeTe_{0.8}S_{0.2}$ and elucidate the component correlated with the superconductivity in $FeTe_{1-x}S_x$. Here we show the effect of soaking in various red wines for superconductivity in $FeTe_{0.8}S_{0.2}$

Experimental methods

The polycrystalline samples of FeTe$_{0.8}$S$_{0.2}$ were synthesized using a solid-state reaction. At first, we synthesized TeS powder as a starting material to avoid evaporation of S during the sintering process. Te and S powders were put into a quartz tube with a nominal composition of Te:S = 1:1. The quartz tube was evacuated by a rotary pump, and then sealed. After being heated at 500 °C for 8 hours, the mixture was ground. Powders of Fe and TeS, and grains of Te with a nominal composition of FeTe$_{0.8}$S$_{0.2}$ were sealed into an evacuated quartz tube, and heated at 600 °C for 10 hours. The obtained mixture was ground, pelletized, and sealed into an evacuated quartz tube. Then the tube was heated again at 600 °C for 10 hours. We prepared six glass bottles filled with several types of red wine made from different grapes. The wine varieties used in this study were Gamay (product name: Beaujolais, year of production: 2009, winery: Paul Beaudet), Merlot (Les Tannes Tradition Merlot, 2010, Jean-Claude Mas), Cabernet Sauvignon (Les Tannes Tradition Cabernet Sauvignon, 2010, Jean-Claude Mas), Pinot Noir (Bourgogne Pinot Noir, 2009, Maison Jean-Philippe Marchand), and Sangiovese (Larinum Sangiovese Daunia, 2009, Caldora s.r.l.). We also prepared Bon Marche (Bon Marche, 2010, Mercian Corporation), which is a blend of several varieties of grapes, as the standard red wine. The alcohol concentration of the Gamay, Merlot, Cabernet Sauvignon, Sangiovese, and Bon Marche were 12.5, 13.5, 13.5, 12.5, 12.5, and 11.0 % respectively. The as-grown FeTe$_{0.8}$S$_{0.2}$ pellet was cut into several pieces, and the pieces were put into each red wine, and were heated at 70 °C for 24 hours. Here we define the sample name as the heating condition; for example, the Bon Marche sample is referred to as the FeTe$_{0.8}$S$_{0.2}$ sample heated in Bon Marche.

The temperature dependence of magnetic susceptibility was measured using a SQUID magnetometer down to 2 K under a magnetic field of 10 Oe. The shielding volume fraction was estimated from the difference of magnetic susceptibility between the value of the normal state and the lowest-temperature value after zero-field cooling.

Results and discussion

Figure 1 shows the temperature dependence of magnetic susceptibility for the as-grown $FeTe_{0.8}S_{0.2}$ sample and the various red wine samples. All the red wine samples show superconductivity, whereas no superconducting signal was observed in the as-grown sample. We estimated the shielding volume fraction of the samples heated in the Gamay, Merlot, Cabernet Sauvignon, Pinot Noir, Sangiovese, and Bon Marche to be 93.9, 82.8, 80.4, 75.2, 71.5, and 61.7 %, respectively.

The obtained shielding volume fractions are summarized in Fig. 2 as a function of alcohol concentration. The previous results reported in ref. 9 are also plotted in Fig 2. For previous results, the shielding volume fractions obtained from alcoholic beverage samples except red wine one were between 23.1 and 46.8 %. On the other hand, the shielding volume fractions obtained from the red wine samples are between 61.7 and 93.9 %: All the red wines are more effective to induce superconductivity in $FeTe_{0.8}S_{0.2}$ than the other alcoholic beverages. From this result, we assume that red wine has a highly effective component correlated with superconductivity in $FeTe_{0.8}S_{0.2}$.

We then performed the component analysis of the red wines to find out the key factor to induce superconductivity using a organic acid analyzer (Prominence, Shimadzu). A lot of components were detected by the component analysis. Among them, we focus on the tartaric acid, because the concentrations of tartaric acid have a positive correlation with the shielding volume fraction obtained from the red wine samples. Figure 3 shows the normalized volume fractions of the red wine samples as a function of concentration of tartaric acid.  Here the shielding volume fraction is normalized by the value of the Bon Marche sample.  We found that the samples with a higher concentration of tartaric acid had a tendency to show a higher superconducting volume fraction.  In order to investigate the effect of tartaric acid for the shielding volume fraction, we prepared aqueous solutions with tartaric acid.  The prepared concentrations of the tartaric acid were 0.82, 2.46, and 4.10 g/L.  Here the tartaric acid was dissolved in water.  The normalized volume fraction obtained from the tartaric acid samples and the water sample are plotted in Figure 4.  The tartaric acid samples show a large volume fraction compared to the water sample.  We confirmed that tartaric acid is a few times more effective in inducing superconductivity in $FeTe_{0.8}S_{0.2}$ than water.  In this study, we found that tartaric acid is one of the key factors to induce superconductivity in $FeTe_{0.8}S_{0.2}$.  However, the values of shielding volume fraction of the samples do not reach that of the sample heated in red wine.  Although the tartaric acid efficiently induces superconductivity in $FeTe_{0.8}S_{0.2}$, it is not as sufficient as red wine.  Therefore, further study to investigate the other key factor inducing superconductivity in $FeTe_{1-x}S_x$ is needed.

Conclusion

  We investigated the red wine dependence of the shielding volume fraction in $FeTe_{0.8}S_{0.2}$ and found that the tartaric acid in red wine is correlated with the superconductivity in $FeTe_{0.8}S_{0.2}$.  We found that tartaric acid is one of the key factors to induce superconductivity in this material, since the volume fraction of the samples heated in red wine is proportional to the concentration of tartaric acid.  However, the values of the shielding volume fraction of the tartaric acid samples do not reach that of the red wine samples. A detailed investigation to clarify the key factor in inducing superconductivity by organic acid is anticipated.

Acknowledgement

This work was partly supported by a Grant-in-Aid for Scientific Research (KAKENHI).

Figure captions

Figure 1 Temperature dependence of magnetic susceptibility for the as-grown sample and the various red wine samples.

Figure 2 The shielding volume fractions of the various red wine samples as a function of alcohol concentration. The previous results reported in ref. 9 are also plotted.

Figure 3 Normalized volume fraction of FeTe$_{0.8}$S$_{0.2}$ samples heated in various alcoholic beverages as a function of the concentration of tartaric acid.

Figure 4 Concentration of tartaric acid dependence of the volume fractions of the sample heated in water and aqueous solutions of tartaric acid with various concentrations.

Figure 1

![Figure 1: ZFC magnetic susceptibility χ(T) − χ(15K) vs Temperature for FeTe$_{0.8}$S$_{0.2}$ annealed at 70°C for 24h in various wines (Bon Marche, Sangiovese, Pinot Noir, Cabernet Sauvignon, Merlot, Gamay) and as-grown, measured at H = 10 Oe.]

Figure 2

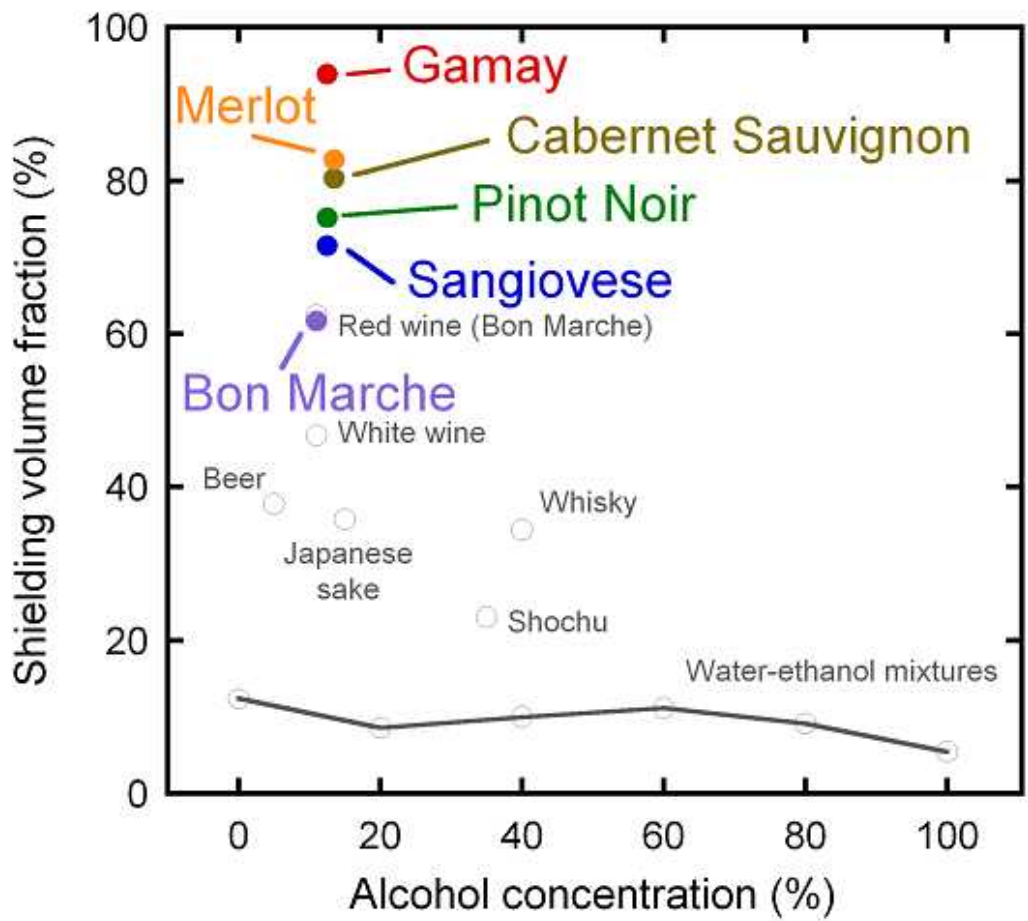

Figure 3

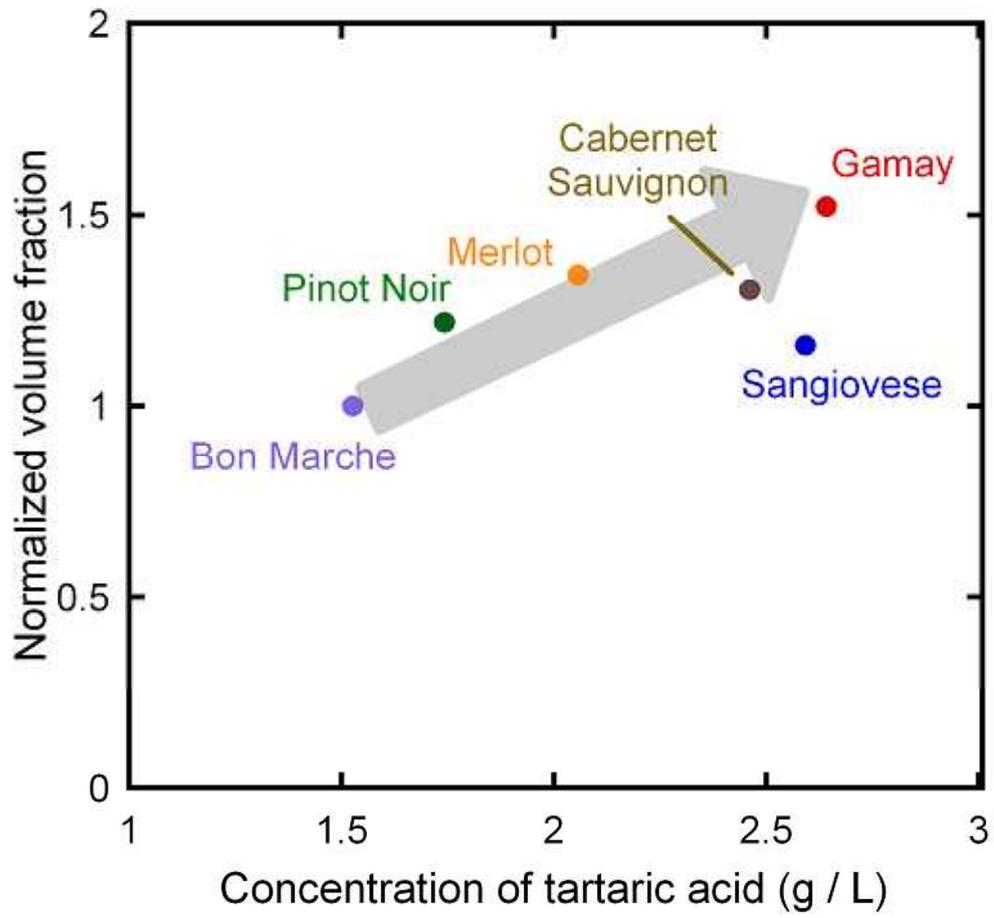

Figure 4

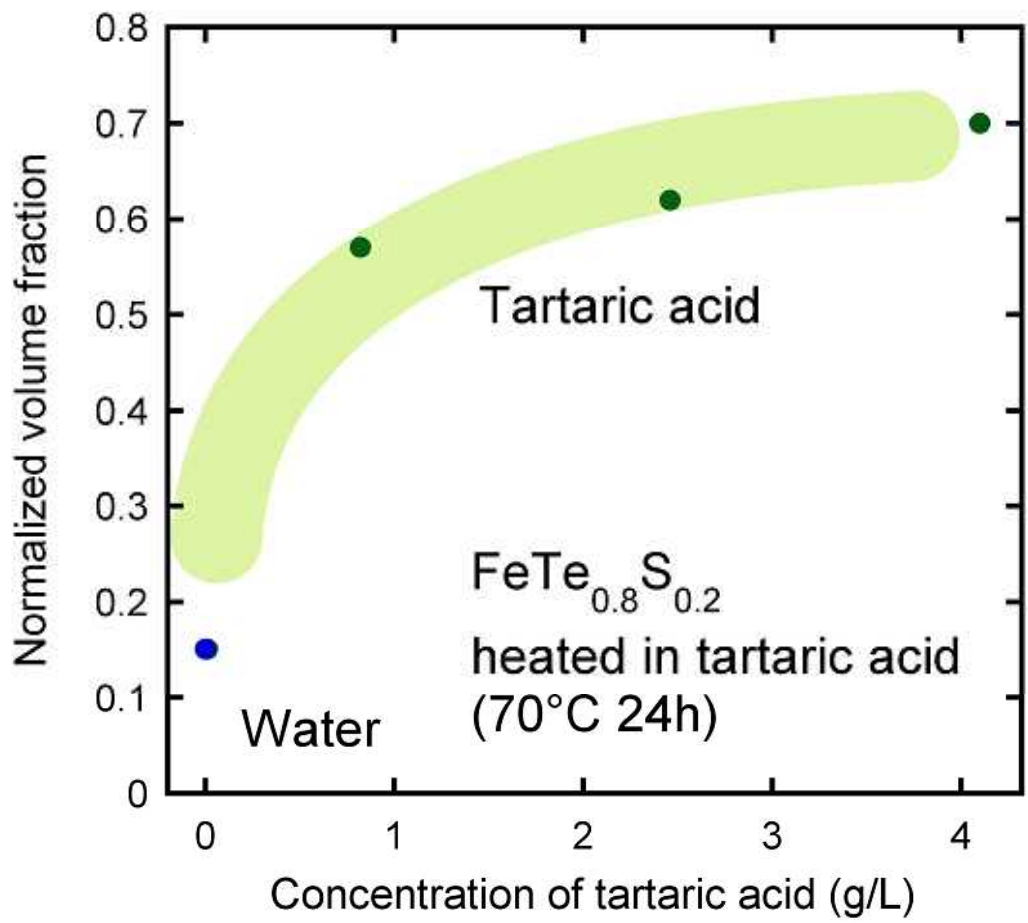